\documentstyle[prb,aps,psfig,multicol]{revtex}
\begin{document} 
\draft 

\title{Magnetic and structural properties of the double-perovskite Ca$_{2}$FeReO$_6$}

\author{W~Westerburg${\dag}$, O~Lang${\ddag}$, C~Felser${\ddag}$, W~Tremel${\ddag}$, M~Waldeck${\S}$, 
        F~Renz${\S}$, P~G\"utlich${\S}$, C~Ritter${\|}$ and G~Jakob${\dag}$\cite{byline}}
\address{${\dag}$Institut f\"ur Physik, Johannes Gutenberg-Universit\"at Mainz, Staudinger Weg 7, 55099 Mainz, Germany}

\address{${\ddag}$Institut f\"ur Anorganische Chemie und Analytische Chemie, Johannes Gutenberg-Universit\"at Mainz, 
Duesbergweg 10-14, 55099 Mainz, Germany}

\address{${\S}$Institut f\"ur Anorganische Chemie und Analytische Chemie, Johannes Gutenberg-Universit\"at Mainz, 
Staudinger Weg 9 , 55099 Mainz, Germany}

\address{${\|}$Institute Laue Langevin, 6, rue Jules Horowitz, Bo\^{\i}te Postale 156, 38042 Grenoble-Cedex 9, France}

\date{17 April 2000} 

\maketitle 

\begin{abstract} 
We suceeded in the preparation of polycrystalline Ca$_{2}$FeReO$_6$ which has a Curie temperature $T_C$ of 540 K,
the highest value of all magnetic perovskites investigated up to now. This material has been characterised by X-ray
and neutron powder diffraction. We found at 548 K a monoclinic unit cell (space group $P2_1/n$) with $a=5.4366(5)$\ \AA, 
$b=5.5393(5)$\ \AA, $c=7.7344(5)$\ \AA{}, and $\beta=90.044(4)^{\circ}$. For low temperatures a phase separation in 
two monoclinic phases with identical cell volume is observed in neutron scattering. The two phases 
possess different magnetic structure and coercivity. 
$^{57}$Fe-M\"ossbauer spectroscopy measurements show the presence 
of four different Fe$^{3+}$ positions indicating two different phases at room temperature, 
indistinguishable in the diffraction experiments. 
The conductivity is thermally activated for all temperatures 
and no significant magnetoresistivity is observed. 

\end{abstract}  



\begin{multicols}{2}
\section{Introduction}
The advent of spin based electronics has lead to a strong interest in materials with high 
spin polarisation. Among these half-metallic oxides are promising candidates for future applications
as magnetoresistive devices. At low temperatures impressive performance of spin
polarised tunneling devices has been reported \cite{Gupta99}. However, the increase of
spin fluctuations with increasing temperature is a severe obstacle for room temperature 
applications of materials with low Curie temperatures. Therefore both high spin polarisation
and high Curie temperatures are important requirements.  
Recently, a large room temperature magnetoresistance was found in Sr$_{2}$FeMoO$_6$
\cite{Kobayashi98}, a material belonging to the class of double-perovskites ($AA'BB'$O$_6$) \cite{Anderson93}
with a Curie temperature above 400~K. 
In this report we present a detailed study of Ca$_{2}$FeReO$_6$ (CFRO), the double-perovskite with the highest 
Curie temperature ($T_C=540$ K). 
\section{Experiment} 
The compound Ca$_{2}$FeReO$_6$ was synthesised by a solid-state reaction from CaO (Alfa, 99.95\%), 
ReO$_3$ (Alfa, 99.9\%), Fe (Alfa, 99.998\%) 
and Fe$_2$O$_3$ (Alfa, 99.99\%) using the stoichiometric ratio of 2:1:$\frac{1}{3}$:$\frac{1}{3}$ 
respectively and batches of an overall mass of 2 g. 
The well grained sample was transferred into a corundum container and sealed in an evacuated quartz tube 
($p=5\times10^{-5}$ mbar).
The tube was heated to 1173 K with a rate of 1 K/min and held at this temperature for 48 hours. After cooling the 
sample to room temperature with a rate of 5 K/min the sample was reground and resealed. The resealed sample was 
annealed at 1173 K for 14 days and then quenched with liquid nitrogen. This method yielded a black polycrystalline 
product.\\
X-ray diffraction was performed at room temperature using a Philips X'Pert MPD diffractometer in Bragg-Brentano
geometry. 
The instrument works with Cu $K_{\alpha}$ radiation ($\lambda = 1.5418$ \AA).\\
At the Institut Laue Langevin in Grenoble the neutron powder diffraction data were collected on the high-resolution
instrument D2B with the sample (10 g) placed in a cylindrical vanadium can inside a cryofurnace. 
A wavelength of $\lambda = 1.594$ \AA{} over an angular range of 0$^\circ$ and 162$^\circ$ was used.
We measured on warming at several fixed temperatures of 2, 100, 200, 300, 400, 444, 524 and 548 K.
The structural and magnetic parameters gained with X-ray and neutron diffraction were refined by the Rietveld method 
using the program {\sc FULLPROF} \cite{Carvajal}. For the line shape a pseudo-Voigt function was selected.
All Bragg peaks could be identified and therefore no regions were excluded in the refinement.\\
The magnetic properties as AC magnetic susceptibility and DC magnetic moment were determined with a Lake Shore 7000
magnetometer and a SHE SQUID magnetometer, respectively.\\
The temperature dependence of the resistivity was measured by the standard four-point technique in a standard
cryostat with a 12 T superconducting magnet.
\section{Crystal structure}
\subsection{Neutron Diffraction}
\begin{figure}[t]
\centerline{\psfig{file=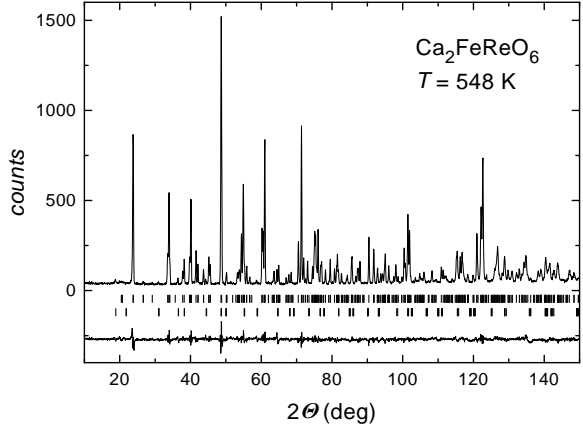,width=0.9\columnwidth}}
\vspace{1.5em} 
\caption{Observed neutron diffraction pattern for CFRO at $T=548$ K above $T_C$.
The difference pattern arises from a refinement including the monolinic CFRO phase and the 
magnetite impurity phase (0.5\%).}
\label{pattern1} 
\end{figure}

\begin{figure}[t]
\centerline{a)\psfig{file=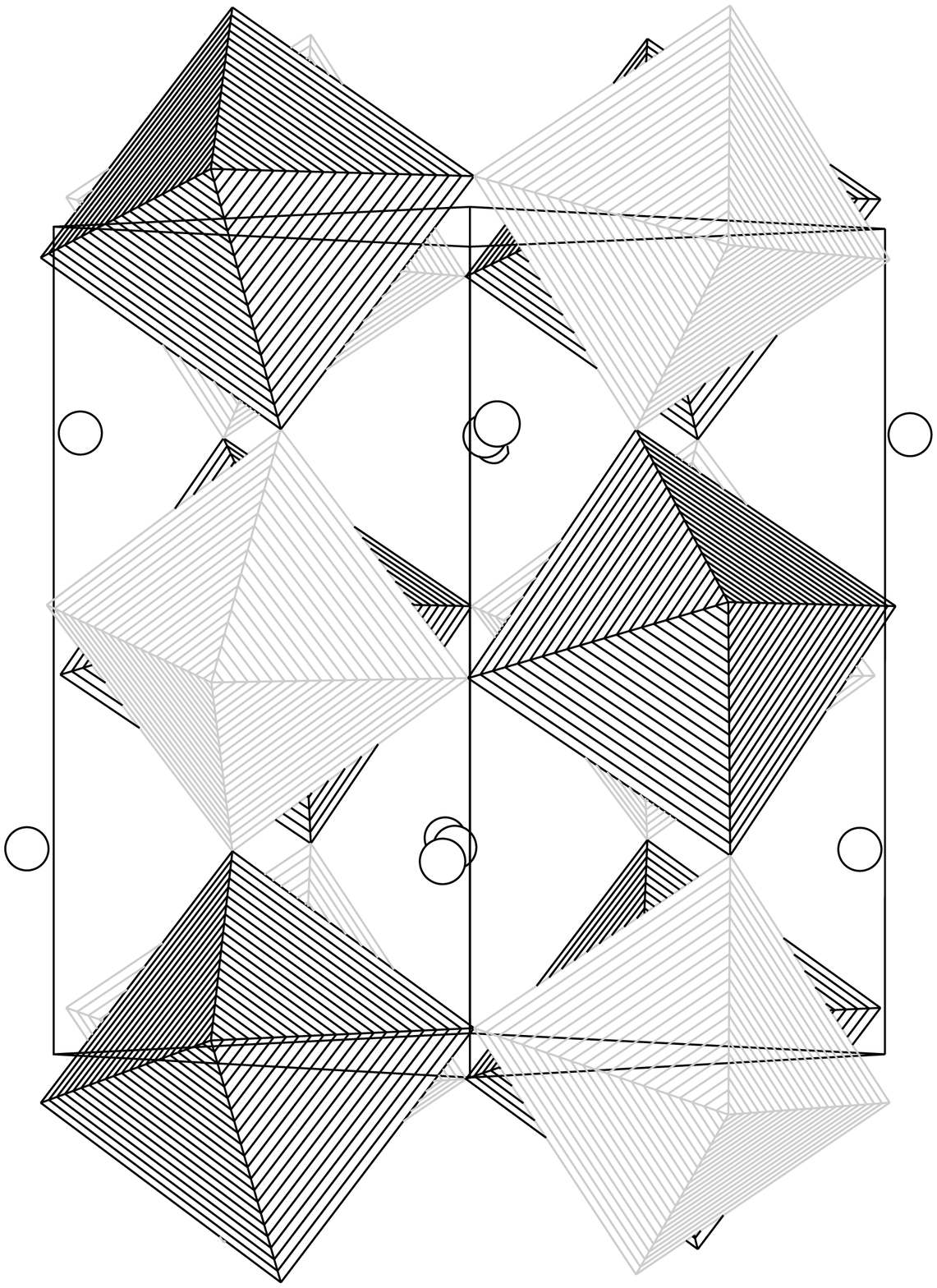,width=0.5\columnwidth}
            b)\psfig{file=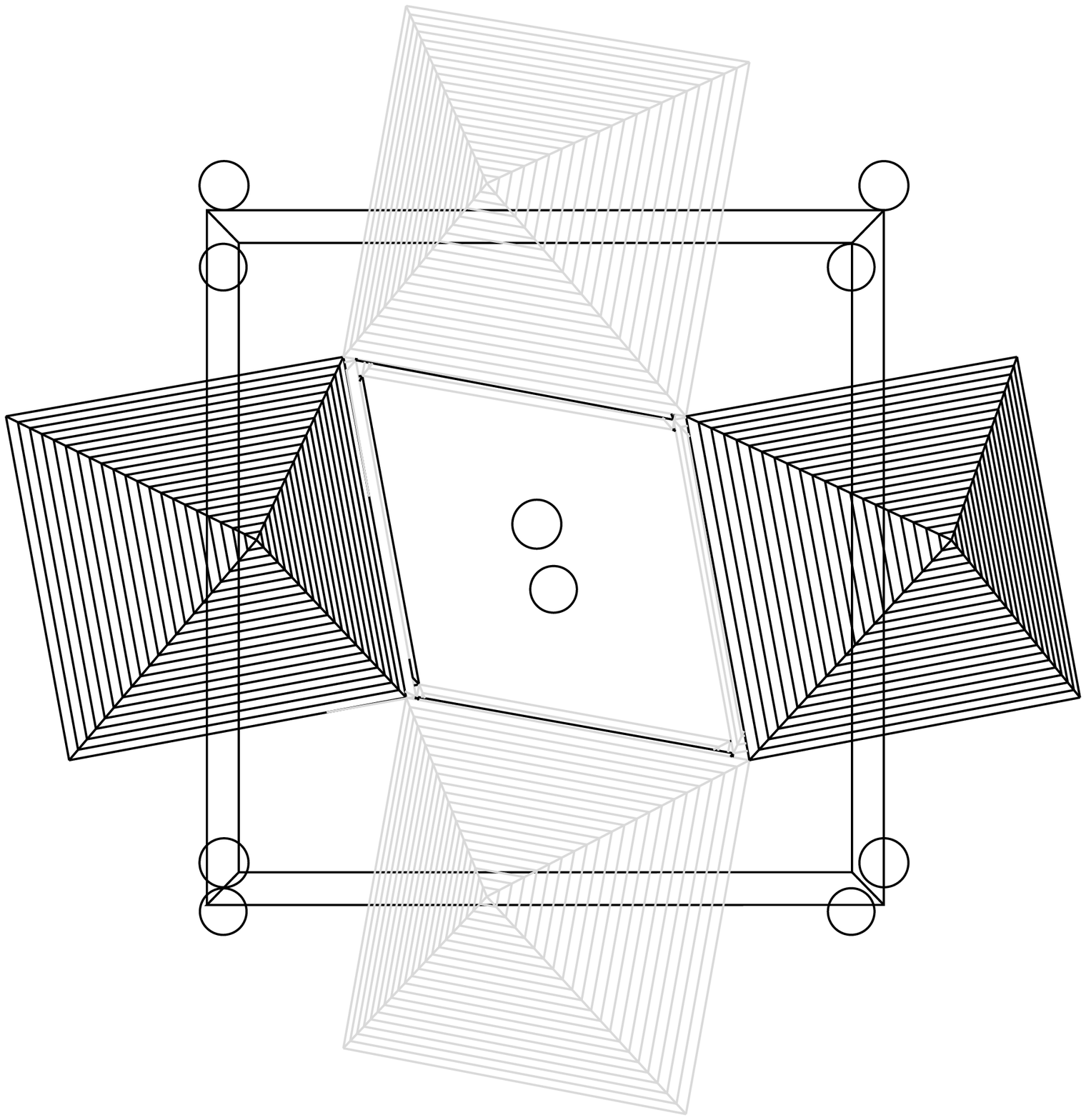,width=0.5\columnwidth}}
\vspace{1.5em} 
\caption{a) View of the unit cell along the crystallographic (110) direction corresponding to
a pseudocubic $a$ or $b$ axis. The monoclinic rock-salt arrangement of the Fe (black) and 
Re (grey) ions with opposite rotations of the octahedra along viewing direction can be seen.\\
b) View along the crystallographic (001) direction showing in phase rotations.}
\label{crystal} 
\end{figure}

Over 300 compounds are known in the class of double-perovskites ($AA'BB'$O$_6$). The $B,B'$-ions arrange
in three different manners, rock-salt, random and in rare cases layered. Which type of configuration exists, depends 
on the charge, size, electronic configuration, and $A$/$B$ size ratio of the involved ions. In the case of 
CFRO the charge difference is 2$e$ (Fe$^{3+}$, Re$^{5+}$, see M\"ossbauer measurements below) 
and the ionic radius difference is 0.065 \AA{}. 
The unit cell for rock-salt arrangement can be derived either from a cubic $2a_0$ or a monoclinic
($\sqrt{2}a_0\times\sqrt{2}a_0\times2a_0$) cell, where $a_0$ is the lattice parameter for the standard cubic 
perovskite $AB$O$_3$ ($a_0\approx$ 4\ \AA). 
The monoclinic cell is favoured if the tolerance
factor $t$ defined in Eq. \ref{tolerance} is less than unity. 
\begin{equation}
t=\frac{\frac{r_A+r_{A'}}{2}+r_O}{\sqrt{2}\left(\frac{r_B+r_{B'}}{2}+r_O\right)}
\label{tolerance}
\end{equation}
The cation order is revealed in the cubic case by the
presence of $(hkl)$ reflections with $h,k,l=2n+1$ or in the monoclinic case 
by $(0kl)$ reflections with $k=2n+1$, respectively. 
For CFRO due to the low tolerance factor of $t=0.89$
a monoclinic unit cell is expected (ionic radii from Ref.\ \cite{Shannon76}). A similar compound
Ca$_{2}$FeMoO$_6$ also a ferrimagnetic oxide with $t=0.88$ was found recently to have a monoclinic
unit cell \cite{Alonso2000}.\\
The diffraction pattern recorded at D2B above $T_C$ at $T=548$ K is shown in Fig.\ \ref{pattern1}.
A slight impurity phase of 0.5\% of Fe$_3$O$_4$ 
(due to reaction 3Fe$_2$O$_3$ $\rightarrow$ 2Fe$_3$O$_4$$ + \frac{1}{2}$O$_2$)
could be detected for all temperatures.  
The pattern was refined in the space group $P2_1/n$. 
The positional and thermal paramaters are listed in Table\ \ref{data3}.
The monoclinic unit cell results from rotations of the
$B$O$_6$, $B'$O$_6$ octahedra. 
Fe and Re are indistinguishable for neutrons in the paramagnetic regime 
due to the small difference in the nuclear coherent scattering lengths of 9.45 fm and 9.20 fm, respectively. 
However, in the ferromagnetic regime they can be discerned due to interaction of the neutron with the 
magnetic moment of the electron shells. Additionally, the structure was checked with X-ray diffraction
(see below).
The refinements show that the Fe atoms occupy the $2d$ position 
$(\frac{1}{2},0,0; 0,\frac{1}{2},\frac{1}{2})$ and Re the $2c$ position 
$(0,\frac{1}{2},0; \frac{1}{2},0,\frac{1}{2})$, i.e.\ there exists an ordered rock-salt arrangement. 
The $A$ atom and three oxygen atoms occupy different $4e$ positions. 
The monoclinic unit cell with tilts of the octahedra is shown in Fig.\ \ref{crystal}. 
According to Glazer's notation we have $a^-a^-b^+$ along the pseudocubic axes \cite{Glazer72}.
The superscripts indicate that neighbouring octahedra along the corresponding axis rotate 
in the same (+) or opposite (-) direction. The view in Fig.\ \ref{crystal}a is along the 
pseudocubic $a$ (or $b$) axis (view along the crystallographic (110) direction)
and shows octahedra rotations with opposite sign. Part b of the 
figure shows the view along the crystallographic $c$ axis showing the in phase rotation of 
the octahedra along this axis. 

\begin{figure}[t]
\centerline{\psfig{file=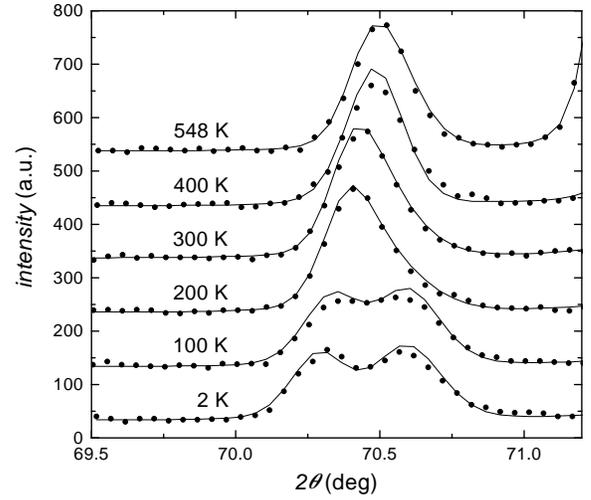,width=0.9\columnwidth}}
\vspace{1.5em} 
\caption{Temperature dependence of the (040) reflection peak of CFRO. With decreasing temperature this peak
with a large momentum transfer along the unique $b$ axis broadens and finally splits into two reflections.
Lines are from refinements including two crystallographic phases as described in the text.}
\label{split} 
\end{figure}

\begin{figure}[t]
\centerline{\psfig{file=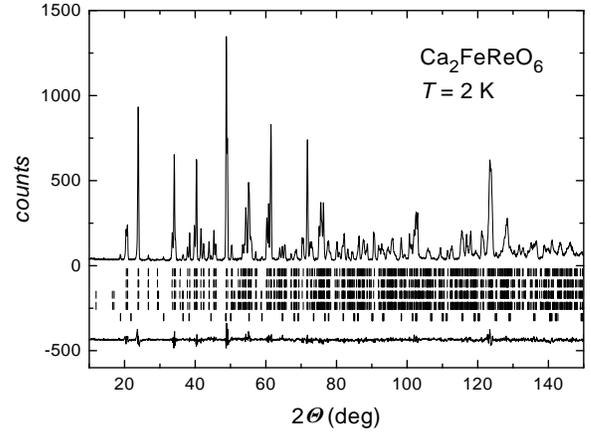,width=0.9\columnwidth}}
\vspace{1.5em} 
\caption{Observed neutron diffraction pattern for CFRO at $T=2$ K.
The difference pattern arises from a refinement including the two magnetic monolinic CFRO phases 
and the magnetite impurity phase (0.5\%).}
\label{pattern2} 
\end{figure}

\begin{figure}[t]
\centerline{\psfig{file=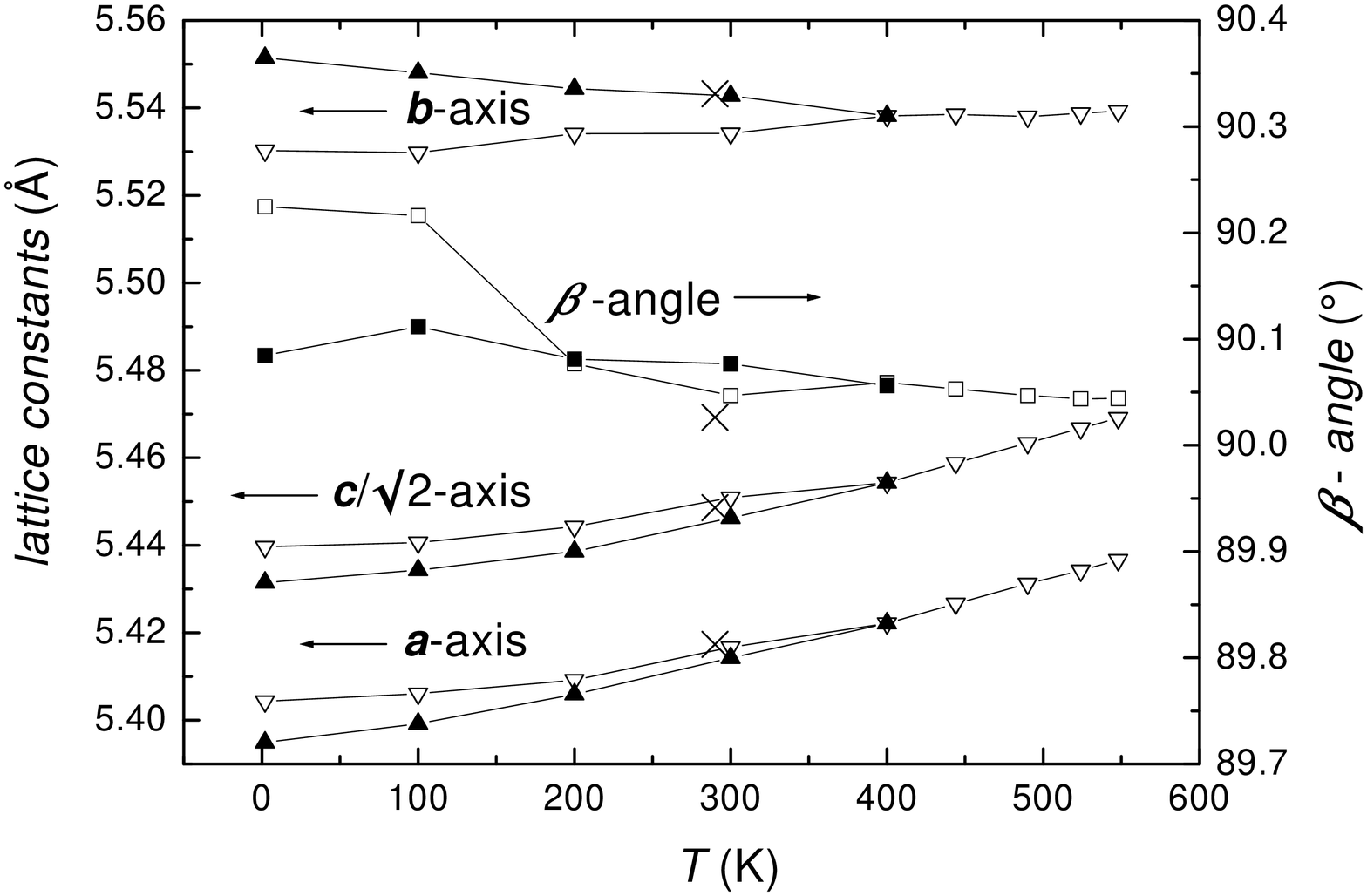,width=0.9\columnwidth}}
\vspace{1.5em} 
\caption{Temperature dependence of the lattice constants (triangles) and the $\beta$-angle (squares) of the two different 
CFRO phases measured by neutron diffraction. The two phases merge to a single phase above room temperature. 
The results from X-ray diffraction 
(only one phase refined) at 290 K are shown by crosses.}
\label{axisvT} 
\end{figure}

With decreasing temperature we observed an unusual peak broadening and finally a peak splitting of 
selective nuclear Bragg reflections with a large momentum transfer along the unique $b$ axis. 
Such an unusual broadening of peaks, which have a large component along the unique axis, was also observed in
perovskite manganites Nd$_{1-x}$Sr$_x$MnO$_3$ which belong to the same space group \cite{Kajimoto99}.
Although the best refinement was achieved in this case using two distinct phases
the origin of the peak broadening has been attributed to the existence of strain effects.
The clear splitting of the (040) reflection visible in the diffraction 
pattern as shown in Fig.\ \ref{split}, however, cannot originate from strain effects 
and clearly two distinct phases are required. In our refinement we used two crystallographic phases 
and two magnetic phases of CFRO and a impurity phase of 0.5\% of magnetite, together five phases. 
The pattern taken at 2 K is shown in Fig.\ \ref{pattern2}.
The two phases of CFRO differ mostly in the values of the $b$ axes
and the $\beta$ angles but have almost the same unit cell volume.
In Table\ \ref{data2} the positional and thermal paramaters of the two phases are shown. 
The results for the lattice constants and the $\beta$ angles of all refinements are listed in Table\ \ref{data1}
and presented in Fig.\ \ref{axisvT}. From 2 K up to 300 K the two phases have almost the same weight in the refinement
but the difference in lattice parameters between the two phases decreases with increasing temperature.
At temperatures of 400 K and higher the $(0k0)$ reflections are symmetric, which is depicted in Fig.\ \ref{split},
and neutron refinement shows a single phase. It is unresolved if the two phases are just similar and can no longer 
be crystallographically distinguished or whether a true phase separation of a single high temperature phase
takes place below 400 K.
The bond length and bond angles are presented in Table\ \ref{data4}
showing again the distorted perovskite structure.
\subsection{X-ray Diffraction}
\begin{figure}[t]
\centerline{\psfig{file=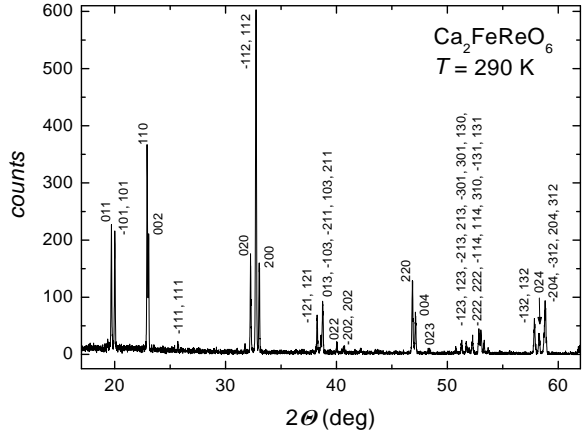,width=0.9\columnwidth}}
\vspace{1.5em} 
\caption{X-ray powder diffraction pattern for CFRO taken at room temperature. The Bragg peaks are indexed
in a monoclinic unit cell $a=5.417(2)$ \AA, $b=5.543(2)$ \AA, 
$c=7.706(2)$ \AA, and $\beta=90.03(3)^{\circ}$ (space group $P2_1/n$).}
\label{xray} 
\end{figure}

The X-ray powder diffraction pattern of CFRO were taken at room temperature. The data are shown in Fig.\ \ref{xray}.
No impurity phase could be detected. Due to the strongly distorted perovskite structure the pattern shows
a large amount of Bragg peaks. The results of the structure refinement are in agreement with the data gained
by neutron diffraction. Due to the high absorption of X-rays in the Re-compound, however, 
the X-ray refinement has larger errors. Nevertheless, the high intensity of the (011) and ($\overline{1}01$), 
(101) reflections indicate the rock-salt arrangement 
of the Fe, Re sublattice and the refinement yields a high degree of order (less than 2\% interchanged Fe, Re atoms).
Due to the lower intensity only one phase was refined. The refined cell parameters of $a=5.417(2)$ \AA, $b=5.543(2)$ \AA, 
$c=7.706(2)$ \AA, and $\beta=90.03(3)^{\circ}$ are shown in Fig.\ \ref{axisvT} by crosses.
\section{Magnetic properties}
For CFRO a Curie temperature of 540 K is reported in literature \cite{Longo61}. However, a detailed investigation
of the magnetic properties is still missing. Therefore we measured AC susceptibility, 
overall and local magnetic moment and magnetic hysteresis.
\subsection{Magnetic susceptibility}
\begin{figure}[t]
\centerline{\psfig{file=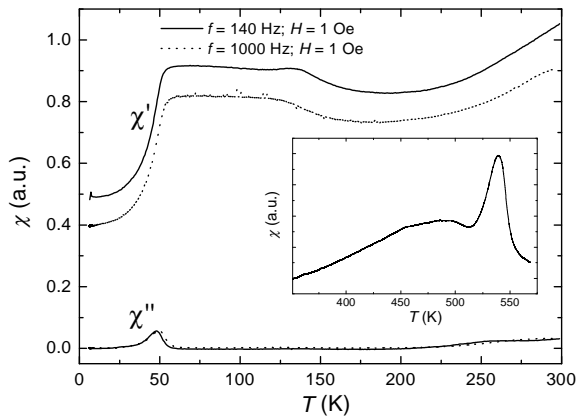,width=0.9\columnwidth}}
\vspace{1.5em} 
\caption{Temperature dependence of the AC susceptibility of CFRO from 4 K up to 300 K. The inset shows the
AC susceptibility above room temperature. Clearly a peak at the Curie temperature of 540 K can be seen.}
\label{acs} 
\end{figure}

In Fig.\ \ref{acs} we show the temperature dependence of the AC susceptibility. Besides the ferro(i)magnetic transition
at 540 K, which is visible in the inset, there are two further anomalies in the AC susceptibility. There 
is a slight temperature dependence of $\chi'$ from room temperature down to 125 K. At this temperature the
susceptibility becomes temperature independent and there is a small anomaly also in the loss component $\chi''$. 
A clear magnetic phase transition exists at 50 K showing up as a sharp decrease of $\chi'$ and a 
sharp maximum in $\chi''$. These anomalies are frequency independent (the measurement taken at 7 Hz not
presented in Fig.\ \ref{acs} shows the
same anomalies) but they do not show up in the temperature dependence of the DC magnetic moment.
\subsection{DC magnetisation}
\begin{figure}[t]
\centerline{\psfig{file=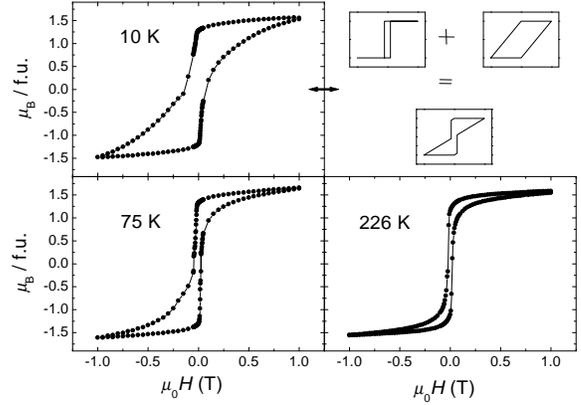,width=0.9\columnwidth}}
\vspace{1.5em} 
\caption{Hysteresis loops of CFRO at constant temperatures of 10 K, 75 K, and 226 K, measured in magnetic
fields from -1 T to 1 T. The unusual shape of the low temperature hysteresis curve results
from two magnetic phases with high and low coercivity, as sketched
in the figure.}
\label{hysteresen} 
\end{figure}

Measurements of the DC magnetisation show ferromagnetic hysteresis loops. A full saturation of the magnetic moment 
at low temperatures is not achieved in fields of $\mu_0H=1$ T.
The unusual shape of the low temperature hysteresis curves can be understood by a superposition of 
two magnetic phases with high and low coercivity, respectively.
They contribute approximately equal 
to the total magnetic moment as is sketched in Fig.\ \ref{hysteresen}. We attribute these two phases to the two different 
crystallographic phases which possess different anisotropy energies. 
With increasing temperature the magnetic phase of high coercitivity becomes 'softer'
and the hysteresis curves of both phases merge to a nearly normal ferromagnetic hysteresis loop.
The remnant magnetic moment of 1.3 $\mu_{\mathbf B}/$f.u. is temperature independent below 250 K while the total coercivity
increases from 11 mT at 250 K to 0.13 T at 4 K.  
\subsection{Magnetic structure from Neutron Scattering}
\begin{figure}[t]
\centerline{\psfig{file=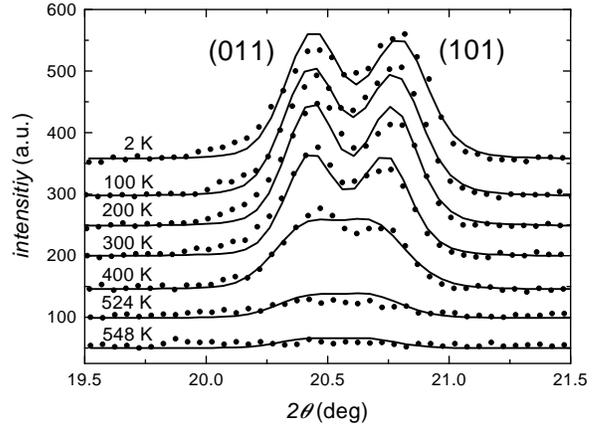,width=0.9\columnwidth}}
\vspace{1.5em} 
\caption{Temperature dependence of the (011) and (101) Bragg peaks which are only visible in
the ferromagnetic regime due to their magnetic origin.}
\label{nuclmagn} 
\end{figure}

Due to the different number of $d$-electrons in the shells Fe and Re are distinguishable magnetically
for neutrons in the ferromagnetic regime.
Large magnetic contributions to the (011) and (101) Bragg peaks are visible around 20.5$^{\circ}$ at low temperatures.
The nuclear contribution to these peaks is negligibly small.
To the best of our knowledge the neutron scattering form factor of Re$^{5+}$ is unknown and we 
approximated it using the values for Mo$^{3+}$ \cite{Wilkinson61}.
Within this approximation the best refinements were obtained with a ferrimagnetic arrangement of the Fe and Re spins
for one phase and a ferromagnetic alignment of the Fe spins for the other phase. 
The magnetic moments at the Fe and Re positions are 4.0(2) and
-0.81(6), respectively for phase 1 and 4.2(2) and -0.1(6), respectively for phase 2.
In both phases best refinements were achieved for spin orientations along the [110] direction.
We obtained at 2 K a Bragg factor of 4.7\% and 4.2\% for the two phases.
The temperature dependence of the (011) and (101) Bragg peaks is shown in Fig.\ \ref{nuclmagn}.
The magnetic contribution decreases with increasing temperature and vanishes at $T_C$.
The magnitude of the refined magnetic moment disagrees with the DC magnetisation result.
In a magnetic field of $\mu_0H=1$ T one should reach in DC magnetisation a value close to saturation, while we only observe
$\approx 1.5 \mu_{\mathbf B}/$f.u.. We do not expect the approximative use of the form factor 
to be responsible for the discrepancy
and since the temperature dependence of the low angle peaks confirms their magnetic origin, 
we consider the neutron result more reliable. Further investigations such as solving the spin-structure
of CFRO are necessary to elucidate this problem.
\section{$^{57}$Fe-M\"ossbauer measurements}
To gain further information of the 
valence and the electronic state of the iron in the compound $^{57}$Fe-M\"ossbauer measurements were performed. 
Natural iron was used for preparation of the sample which contains only
2.2\% $^{57}$Fe.
The presence of Re in the sample caused some difficulties
because of its strong absorption for $\gamma$-rays.
The L-I, L-II, and L-III absorption edges of Re have an energy
slightly below the 14.4\ keV $\gamma$-quanta, 
which are used for $^{57}$Fe-M\"ossbauer measurements. 
Therefore the resulting spectra have a low relative transmission and are 
difficult to fit.

M\"ossbauer spectra of CFRO were recorded at 4.2~K and 293~K
in two different apparatures.
Each spectrum has been taken for one week.
M\"ossbauer transmission spectra were taken with a constant acceleration
spectrometer, using a 1024 channel analyser in the time mode.
For the 4.2~K measurement a Na(Tl)I scintillator was used
while the 293~K measurement was done with a proportional counter.
A 50~mCi $^{57}$Co source in a rhodium matrix at ambient temperature was used
in both cases.
The spectra were fitted with the M\"ossbauer analysis program 
\emph{effi} \cite{GERD2000}.

\begin{figure}[t]
\centerline{\psfig{file=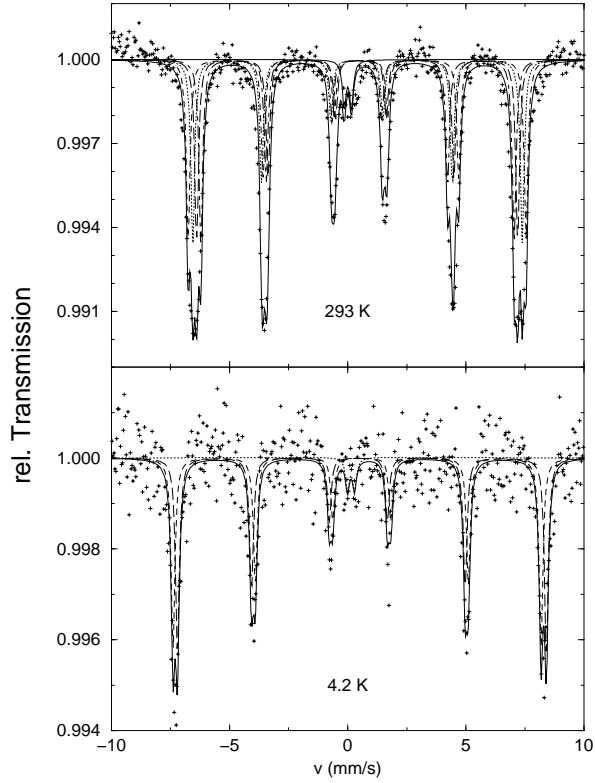,width=0.9\columnwidth}}
\vspace{1.5em} 
\caption{$^{57}$Fe-M\"ossbauer spectra of CFRO recorded at 4.2 K and 293 K.
The total fit curve (solid line) is composed of a
super-positioning of the lines for phase 1 (dotted and dotted-dashed) 
and for phase 2 (dashed and long dashed).}
\label{CaFeRe_MB}
\end{figure}

$^{57}$Fe-M\"ossbauer spectra were used to resolve the valence 
and electronic state of the iron in the compound.
The M\"ossbauer data are shown in Fig.\ \ref{CaFeRe_MB} and 
Table\ \ref{CaFeRe_MB_TAB}.
For both temperatures a six-line pattern, 
which is typical for magnetically long range ordered systems \cite{GREE1971}, 
was obtained.
At 293\ K, M\"ossbauer spectroscopy revealed four iron sites 
which is understood, knowing the presence of two phases from neutron diffraction measurements 
with two different Fe sites for each phase.
Both spectra were fitted with natural line width.
In each phase, the isomer shift $\delta$ and the quadrupole splitting $\Delta E_{\mathrm Q}$ were taken
correlated, while the magnetic splitting $B_{\mathrm hf}$ appeared to be different for all four 
sites.
At 4.2\ K, the difference in the magnetic splittings diminished so that two
sites in each phase appear almost indistinguishable.
Such a behaviour is typical for magnetic systems approaching magnetic saturation.

The subspectrum with an isomer shift $\delta$~=~0.004(5)\ mm/s at ambient 
temperature and $\delta$~=~0.139(6)\ mm/s at 4.2\ K results from an 
Fe$^{3+}$-impurity in the Be-window of the proportional counter.

The hyperfine fields are in the same range as those for the analogous Ba 
compound Ba$_2$FeReO$_6$ measured by Sleight and Weiher \cite{SLEI1974}.
In the recent study on Ca$_2$FeMoO$_6$ the values for the magnetic fields
are almost the same \cite{PINS1999}.
There they found at 4.2 K two iron sites in the ratio 0.44 : 0.56. 
Unfortunately no satisfying conclusion can be drawn on whether 
the irons are in two phases with one site or in one phase with two sites.

In CFRO the value for quadrupole splittings is practically temperature
independent, 
which is typical for Fe$^{3+}$ high spin in octahedral site \cite{GUET1978}.
Also the shifts, quadrupole splittings, and hyperfine fields
are in the range of a typical Fe$^{3+}$ in the high-spin state
in nearly octahedral environment \cite{GREE1971}.
\section{Transport properties}
\begin{figure}[t]
\centerline{\psfig{file=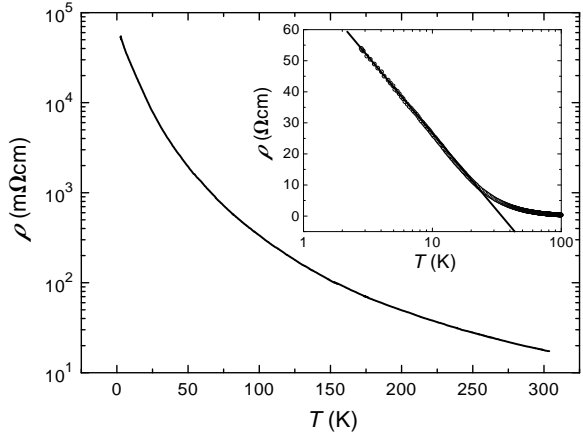,width=0.9\columnwidth}}
\vspace{1.5em} 
\caption{Temperature dependence of the longitudinal resistivity. Above 110 K the resistivity is described
by a variable range hopping like model. Below 20 K the resisitivity increases logarithmic with falling
temperature as can be seen in the inset.}
\label{RvT} 
\end{figure}

In Fig.\ \ref{RvT} the longitudinal resistivity of CFRO is shown. The room temperature value of 
17 m$\Omega$cm is comparable with other reported results \cite{Prellier99}. The temperature dependence shows a
thermally activated behaviour. Down to 4\ K the resistivity increases more than three orders of magnitude
to 53 $\Omega$cm at 2.7\ K.
Even in magnetic fields of $\mu_0H=8$ T no magnetoresistance was observed over the whole temperature range.
This is in contrast to the similar compound Ca$_2$FeMoO$_6$ which shows a metallic behaviour and a 
large magnetoresistance \cite{Alonso2000}.
Closer inspection of data reveals a change in conduction mechanism. 
Below 20 K the resistivity increases strictly logarithmic with decreasing temperature, as shown 
in the inset of Fig.\ \ref{RvT}. 
Above 110 K the temperature dependence of the resistivity is variable range hopping like,
$\rho \propto \exp((T_0/T)^{0.25})$.
A similar behaviour of the resistivity was found in disordered Sr$_2$FeMoO$_6$ thin films \cite{wester2000MMM}.
\section{Conclusions}
In summary, we have investigated the crystal structure, the transport and magnetic properties of polycrystalline
double-perovskite Ca$_{2}$FeReO$_6$. 
We found a monoclinic unit cell with rock-salt order of the Fe and Re ions.
Below $T_C=540$ K the material 
is magnetically ordered. By M\"ossbauer measurements a typical Fe$^{3+}$ state was revealed.
For low temperatures a phase separation in 
two monoclinic phases with identical cell volume is observed in neutron scattering. The two phases 
possess different magnetic structure and coercivity. 
The temperature dependence of the resistivity exhibits a thermally activated behaviour
and shows no magnetoresistance over the whole temperature range. 
Diffraction measurements on a single crystal are necessary to identify the origin of the phase separation.
\acknowledgments
This work was supported by the Deutsche Forschungsgemeinschaft through Project JA821/1-3, Gu95/47-2 and the 
Materialwissenschaftlichen Forschungszentrum (MWFZ) Mainz. The ILL laboratory is acknowledged for
granting beam time. We thank J. Ensling, H. Spiering, and H.J. Elmers for useful discussions about 
M\"ossbauer and magnetisation measurements.

\begin{table}[t]
\caption{Positional and thermal paramaters of CFRO (space group $P2_1/n$) at 548 K.
The $R$-factor was 4.4\%.}
\vspace{1.5em} 
\begin{tabular}{cccccc}
Atom & Site & $x$ & $y$ & $z$ & $B$ \\ 
 & & & & & (\AA$^2$) \\ \hline
		                  Ca & $4e$ & 0.0128(7)     & 0.0432(7)     & 0.7484(7) & 1.11(8) \\
	        	          Fe & $2d$ & $\frac{1}{2}$ & 0             & 0         & 0.63(5) \\
     		        	  Re & $2c$ & 0             & $\frac{1}{2}$ & 0         & 0.15(3) \\
			   	  O1 & $4e$ & 0.2914(7)     & 0.2940(7)     & 0.9591(7) & 0.69(7) \\ 
			          O2 & $4e$ & 0.2996(7)     & 0.2918(7)     & 0.5389(7) & 1.22(8) \\
 			          O3 & $4e$ & 0.9206(7)     & 0.4791(7)     & 0.7514(7) & 0.94(7) \\ \hline
\end{tabular}
\label{data3}
\end{table}

\begin{table}[t]
\caption{Positional and thermal paramaters of CFRO (space group $P2_1/n$) at 2 K.
The $R$-factors for the two phases were 4.1\% and 3.6\%, respectively.}
\vspace{1.5em} 
\begin{tabular}{ccccccc}
Phase & Atom & Site & $x$ & $y$ & $z$ & $B$ \\ 
& & & & & & (\AA$^2$) \\ \hline
		                 & Ca & $4e$ & 0.0078(5)     & 0.0482(5)     & 0.7485(5) & 0.14(5) \\
	        	         & Fe & $2d$ & $\frac{1}{2}$ & 0             & 0         & 0.05(3) \\
     		        	 & Re & $2c$ & 0             & $\frac{1}{2}$ & 0         & 0.00(1) \\
\raisebox{3ex}[3ex]{\bf 1}    	 & O1 & $4e$ & 0.2931(5)     & 0.2928(5)     & 0.9564(4) & 0.45(4) \\ 
			         & O2 & $4e$ & 0.3025(5)     & 0.2951(5)     & 0.5414(4) & 0.37(4) \\
 			         & O3 & $4e$ & 0.9148(5)     & 0.4748(5)     & 0.7516(4) & 0.35(4) \\ \hline
		                 & Ca & $4e$ & 0.0158(5)     & 0.0503(5)     & 0.7517(4) & 0.22(5) \\
	        	         & Fe & $2d$ & $\frac{1}{2}$ & 0             & 0         & 0.06(3) \\
     		        	 & Re & $2c$ & 0             & $\frac{1}{2}$ & 0         & 0.00(1) \\
\raisebox{3ex}[3ex]{\bf 2}   	 & O1 & $4e$ & 0.2990(5)     & 0.2993(5)     & 0.9531(4) & 0.33(4) \\ 
			         & O2 & $4e$ & 0.2944(5)     & 0.2917(5)     & 0.5439(4) & 0.38(4) \\
 			         & O3 & $4e$ & 0.9171(5)     & 0.4749(5)     & 0.7516(4) & 0.35(4) \\
\end{tabular}
\label{data2}
\end{table}
\end{multicols}
\begin{table}[t]
\caption{Lattice constants and $\beta$ angles as a function of temperature of the two phases. Above 300 K the
two phases merge to a single phase.}
\vspace{1.5em} 
\begin{tabular}{ccccccccc}
        & \multicolumn{4}{c}{Phase 1} & \multicolumn{4}{c}{Phase 2} \\ \hline
$T$  & $a$ & $b$ & $c$ & $\beta{}$ & $a$ & $b$ & $c$ & $\beta{}$ \\ 
(K)  &(\AA)&(\AA)&(\AA)&(deg)      &(\AA)&(\AA)&(\AA)&(deg) \\ \hline
  2  & 5.4043(5) & 5.5303(7) & 7.6923(5) & 90.225(4) & 5.3949(5) & 5.5515(7) & 7.6812(5) & 90.085(4) \\
100  & 5.4060(5) & 5.5297(7) & 7.6942(5) & 90.216(4) & 5.3992(5) & 5.5480(7) & 7.6853(5) & 90.112(4) \\
200  & 5.4091(5) & 5.5341(7) & 7.6997(5) & 90.077(4) & 5.4059(5) & 5.5444(7) & 7.6913(5) & 90.081(4) \\
300  & 5.4167(5) & 5.5342(7) & 7.7091(5) & 90.047(4) & 5.4142(5) & 5.5428(7) & 7.7021(5) & 90.077(4) \\ \hline
400  & 5.4222(5) & 5.5382(5) & 7.7136(5) & 90.059(4) &-&-&-&-\\
444  & 5.4267(5) & 5.5385(5) & 7.7199(5) & 90.053(4) &-&-&-&-\\
490  & 5.4312(5) & 5.5380(5) & 7.7264(5) & 90.047(4) &-&-&-&-\\
524  & 5.4342(5) & 5.5387(5) & 7.7311(5) & 90.044(4) &-&-&-&-\\
548  & 5.4366(5) & 5.5393(5) & 7.7344(5) & 90.044(4) &-&-&-&-\\
\end{tabular}
\label{data1}
\end{table}
\begin{multicols}{2}
\begin{table}[t]
\caption{Bond lengths and bond angles for CFRO at 2 K and 548 K.}
\vspace{1.5em} 
\begin{tabular}{cccc}
Temperature  & \multicolumn{2}{c}{2 K}       & 548 K  \\ \hline
Phase   & {\bf 1}      & {\bf 2}      &       \\ \hline
\multicolumn{4}{c}{Main bond lengths of FeO$_6$ octahedra (\AA)} \\
Fe---O1   & 1.9955 & 2.0163             & 2.0094  \\
Fe---O2   & 2.0134 & 1.9930             & 2.0181  \\
Fe---O3   & 1.9960 & 1.9892             & 1.9954  \\
\multicolumn{4}{c}{Main bond lengths of ReO$_6$ octahedra (\AA)} \\
Re---O1   & 1.9846 & 1.9937             & 1.9781  \\
Re---O2   & 1.9765 & 1.9918             & 1.9725  \\
Re---O3   & 1.9686 & 1.9640             & 1.9737  \\
\multicolumn{4}{c}{Bond angles (deg)} \\
Fe---O1---Re ($\times{}2$) & 152.524 & 149.691 & 153.423 \\        	          
Fe---O2---Re ($\times{}2$) & 151.383 & 152.490 & 153.046 \\        	          
Fe---O3---Re ($\times{}2$) & 151.909 & 152.581 & 153.970 \\
\multicolumn{4}{c}{Short bond lengths Ca---O (\AA)} \\
Ca---O1                    & 2.3770  & 2.3279  & 2.3700  \\        	          
Ca---O1                    & 2.5972  & 2.5751  & 2.6222  \\        	          
Ca---O1                    & 2.6735  & 2.7052  & 2.6953  \\        	          
Ca---O2                    & 2.3684  & 2.3600  & 2.3840  \\        	          
Ca---O2                    & 2.6379  & 2.5710  & 2.6376  \\        	          
Ca---O2                    & 2.6604  & 2.6910  & 2.6891  \\        	          
Ca---O3                    & 2.3196  & 2.3728  & 2.3828  \\        	          
Ca---O3                    & 2.4123  & 2.4166  & 2.4662  \\        	          
\end{tabular}
\label{data4}
\end{table}

\begin{table}[t]
\caption{M\"ossbauer parameters, $\delta$: isomer shift, $B_{\mathrm hf}$: hyperfine field, and
$\Delta E_{\mathrm Q}$: quadrupole splitting.}
\vspace{1.5cm}
\begin{tabular}{cccccc}
$T$ & Iron Site &$\delta$&$B_{\mathrm hf}$&$\Delta E_{\mathrm Q}$&Ratio\\
(K)&&(mm/s)&(T)&(mm/s)&\\ \hline
&1 a&0.477(6)$^a$&44.32(5)&$-$0.16(1)$^a$&0.494(5)$^a$\\
&1 b&0.477(6)&42.11(5)&$-$0.16(1)&0.494(5)\\
\raisebox{3ex}[3ex]{293}&2 a&0.401(5)$^b$&43.22(5)&$+$0.03(1)$^b$&0.506(5)$^b$\\
&2 b&0.401(5)&41.10(5)&$+$0.03(1)&0.506(5)\\\hline
&1&0.495(7)&48.27(4)&$-$0.20(1)&0.494(5)$^c$\\
\raisebox{3ex}[3ex]{4.2}&2&0.524(6)&48.39(4)&$+$0.15(1)&0.506(5)$^c$\\ 
\end{tabular}
\label{CaFeRe_MB_TAB}
$^a$correlated with iron site 1 b, $^b$correlated with iron site 2 b,
$^c$ratio taken from ambient temperature spectrum
\end{table}
\end{multicols}
\end{document}